\documentclass[12pt]{JHEP3}
\usepackage{amsfonts,amssymb}
\usepackage{amsmath, young}

\usepackage{epsf}

\def\T11{{T}^{1,1}}
\def\bear{\begin{eqnarray}}
\def\eear{\end{eqnarray}}

\newcommand{\vac}{{|0\rangle}}

\newcommand{\pa}{\partial}
\newcommand{\tr}{{\rm tr}}
\newcommand{\comment}[1]{}
\newcommand{\pasl}{\pa\kern-.55em /}

\newcommand{\ksl}{k\kern-.55em /}

\DeclareFixedFont{\xiiss}{OT1}{cmss}{m}{n}{12}
\DeclareFixedFont{\ixss}{OT1}{cmss}{m}{n}{9}
\DeclareFixedFont{\cmrnine}{OT1}{cmr}{m}{n}{9}
\newcommand{\field}[1]{\mathbb{#1}}

\newcommand{\BC}{{\field C}}
\newcommand{\BR}{{\field R}}
\newcommand{\BZ}{{\field Z}}

\newcommand{\CCs}{\hbox{\ixss C\kern-.4emI}}
\newcommand{\ZZs}{\hbox{\ixss Z\kern-.4emZ}}

\newcommand{\diag}{\hbox{diag}}

\title{A matrix model for a quantum hall droplet with manifest particle-hole symmetry.}
\author{David Berenstein\\ 
Department of Physics, UCSB, Santa Barbara, CA 93106\\
Email: \email{dberens@physics.ucsb.edu}}
\abstract{We find that a gauged matrix model of rectangular fermionic matrices (a matrix version of the fermion harmonic oscillator)
realizes a quantum hall droplet with manifest particle-hole symmetry. The droplet consists
of free fermions on the topology of a sphere. It is also possible to deform the Hamiltonian by double trace operators, and we argue that this device can produce two body potentials which might lead the system to realize a fractional quantum hall state on the sphere. We also argue that a single gauged fermionic quantum mechanics of hermitian matrices realizes a droplet with an edge that has $c=1/2$ CFT on it.}

\keywords{Matrix models, Quantum hall effect}
\preprint{hep-th/0409115}

\begin{document}

\section{Introduction}

Calculating the quantum levels of a non-relativistic electron in a 
uniform magnetic field (in two dimensions) is a well known problem in quantum mechanics. 
From the point of view of the phase space of the classical system, with coordinates $q_1,p_1,q_2,p_2$, the Hamiltonian  is a sum of two squares, and this is formally given by a quadratic form of rank two, namely
\begin{equation}
H = \mu \beta_1^2 + \rho \beta_2^2
\end{equation}
The classical poisson bracket of $\beta_1$ and $\beta_2$ is non-zero. In this sense, this is a generic quadratic semi-positive form of rank two on the phase space. These two coordinates can be considered as canonical conjugates of each other.
 One can easily find two linear combinations of the form 
\begin{eqnarray}\tilde q_1&=& q_1+a_1 \beta_1 +b_1\beta_2\\
\tilde q_2 &=& q_2+a_2\beta_1+b_2\beta_2
\end{eqnarray}
such that their Poisson brackets with $\beta_1$ and $\beta_2$ are zero, and thus they are constants of motion under the Hamiltonian evolution. Classically, these two deformed  coordinates also have a non-zero Poisson bracket. 
The classical motion in the magnetic field is given by circular orbits, which are centered at $\tilde q_1$ and $\tilde q_2$.

When we turn the problem to quantum mechanics, the two conjugate coordinates $\beta_1$ and $\beta_2$ determine a single harmonic oscillator. The levels of this harmonic oscillator are the Landau levels. Now, $\tilde q_1$ and $\tilde q_2$ are non-commuting operators, which parametrize the degeneracy of each Landau level. They can also be described as the phase space of a single classical variable (for the sake of argument it can be $\tilde q_1$). 
Since these two coordinates don't commute, there is a minimum area that the wave function occupies in the phase space determined by $\tilde q_1, \tilde q_2$,  which can be identified with
$\hbar$ after rescalings. A basis of states that realizes the uncertainty bound and 
which are localized in both $\tilde q_1$ and 
$\tilde q_2$ can be given by coherent states, where we are allowed to talk about values for $\tilde q_1$ and $\tilde q_2$ where a wave function is centered.

When we consider a system of free fermions (lets say $k$), the lowest lying state of the system 
will have all the fermions in the lowest Landau level, but at different values of $\tilde q_1$ and $\tilde q_2$. Thus, to describe the low energy physics of the system (the degenerate vacua)
we can forget completely the coordinates $\beta_1$ and $\beta_2$ which describe the higher Landau levels, and set the system to the vacuum for these two coordinates.

 In order to count states, we can break the degeneracy of the system by introducing a small perturbing potential which is a function of $\tilde q_1$ and $\tilde q_2$ only, which for simplicity we take to be quadratic in $\tilde q_1$ and $\tilde q_2$, but with very small coefficients, so that the energy differences associated to the frequency of the $\tilde q_1$ and $\tilde q_2$ oscillator are much smaller than the ones associated to $\beta_1, \beta_2$, as well as the Fermi level. We choose the function so that the level sets of the hamiltonian are compact. We can also choose it so that the level sets are given by hyperbolas, and that corresponds to the $c=1$ matrix model (see \cite{Kleb} for an introductory review).
 
Fermi statistics will force the particles to be located at different values of $\tilde q_1$ and $\tilde q_2$. With our choice for potential for the $\tilde q_i$, the Fermi surface will be a circle in the $(\tilde q_1, \tilde q_2)$ plane, and the fermions form a droplet of constant density in this plane.
This is the quantum hall droplet (see for example \cite{QHE}). 
The fact that this droplet is described in a phase space of a single classical variable has been used to provide some hydrodynamical description in terms 
of a non commutative $U(1)$ gauge theory \cite{Suss}.

There are various ways to describe the excitations of the system. One can consider
the collective motion of particles that deforms the Fermi surface, and one can also consider
taking individual particles from the top of the Fermi surface and giving them a lot of energy.
The deformations of the Fermi surface are given by a free (chiral) boson in $1+1$ dimensions, and this is an 
example of bosonization (see \cite{Stoneedge,IKS,IKS2} for a description of the edge physics).
One can also create collective excitations of the particles that describe holes in the Fermi surface. When we are near the top of the Fermi sea, particle and hole states behave very similar, so one has a symmetry that can exchange particle states and hole states.
This symmetry only appears after quantization. Classically, one has no holes, and the particles occupy zero area. This symmetry is a property of the quantized system, but it is not a property of the classical dynamics of the theory.
In every event that one has a symmetry in a quantum system, one would like to describe the system in such a way that the symmetry is manifest, so that one does not have to solve the system before seeing if it is there or not.

There is a second route that eventually leads to the same system. This is to consider a gauged
$U(N)$ matrix model for a Hermitian matrix $X$, with the $U(N)$ group action by conjugation on $X$, and whose action is
\begin{equation}
S = \frac 12 \int dt \tr ((D_t X)^2-X^2) 
\end{equation}
The system classically allows a separation of variables into the eigenvalues of $X$, which we label by $\lambda$.
Quantum mechanically this is still true, and due to measure factors the eigenvalues 
are Fermions \cite{BIPZ}. Each such eigenvalue has a Hamiltonian which is given by
\begin{equation}
H = \frac 12 \int dt \tr (p_\lambda^2+\lambda^2) 
\end{equation}
so we end up in the same system that we described above.

The wave functions of the $\lambda$ can be described in terms of a complete basis of symmetric polynomials in the $\lambda$ times the VanderMonde determinant of the 
$\lambda$ times an universal factor $\exp(-\lambda^2/2)$.
There are various choices of such symmetric polynomials. One which is very easy to write down is the set of all polynomials in the sums of the powers of $\lambda$. 
Each such sum over $\lambda$ is of trace 
form $\sum_\lambda\lambda^n \sim \tr(X^n)$. A less obvious basis is given in terms of Schur polynomials for the $\lambda$\cite{Stschur}. These can be related to characters of $X$ in various representations of $U(N)$, which is a very natural idea in the matrix model.
This is analogous to Wilson lines in various representations of the gauge group in 2D QCD ( see \cite{Gross,GT,MP}).
These actually offer very simple descriptions of particles (as characters of completely symmetric representations of $U(N)$)
and holes (as completely antisymmetric representations of $U(N)$ \cite{Stschur,Poly}). At 
this level, the particle--hole symmetry corresponds to changing symmetry types of representations.

From the point of view of the Young tableaux that characterize the representations of 
 $U(N)$, this corresponds to performing a mirror image of the tableaux along the diagonal
 (we will refer to this as flipping the tableaux). 
 
The symmetry under permutations of the eigenvalues (the statistics of the particles) is embedded in the $U(N)$ group, and it is the residual gauge symmetry after we have chosen $X$ to be diagonal. In this system, we have a finite droplet of quantum hall liquid in an infinite sea of holes. 
If we want a system that describes finitely many particles and holes, so that hey can appear symmetrically, we would like to see the statistics under permutation of particle wave functions particles and the statistics of hole wave functions in the same way as above: embedded in the gauge group. This suggests having a theory with a $U(N)\times U(M)$ symmetry, which after taking eigenvalues appropriately
reduces to a $S_N\times S_M$ symmetry of permutations of eigenvalues.
The easiest way to connect the $U(N)$ and $U(M)$ symmetries by a matrix, is to consider a
pair of rectangular matrices $X, Y$ that transform in the $(N,\bar M)$ and $(\bar N, M)$
representation of the group. We can make the system `Hermitian' if we consider $X=Y^\dagger$, and then we can consider the gauged matrix quantum mechanics of $X,Y$ with a simple quadratic action. 
Such constructions have appeared in studies of the $c=1$ matrix model, when one considers orbifolds of the matrix model in the dual non-critical string theory \cite{DKKMMS}.

However, the idea of the particle hole like symmetry that 
would exchange $M\leftrightarrow N$ does not work for that system, the first one that one would consider.  However, if we let $X,Y$ be fermionic oscillators instead of bosons, then the symmetry under the exchange $M\leftrightarrow N$ exchanges particles and holes, once these states have been identified in the theory. This second option can be also suggested by
realizing that if we have only finitely many holes and particles, then the total phase space 
should have area $N+M$ and be compact. Thus the quantum system should only have a finite dimensional Hilbert space describing the system. This is one feature that the fermionic matrix model realizes automatically. Droplets of quantum hall liquids  on various Riemann surfaces
and higher dimensional spaces have been considered recently in \cite{ZH,KN,Poly2}.

This paper describes how this symmetry can be understood as particle-hole symmetry in detail for the above system. The paper is organized as follows.
In section \ref{sec:matrixQHE} we describe the relationship between
 matrix models and the quantum hall effect in detail. This is review material.
Next, in section \ref{sec:0A} we describe the system of Hermitian rectangular
bosonic  matrices, which we label $0A$ harmonic oscillator to follow the conventions from string theory. Some of these results are probably not new, but I am not aware of work where this is described in he way I present it.
Here we pay special attention to an $SL(2,\BR)$ symmetry of the system, of which the Hamiltonian is one of the generators. We show that single particle states are uniquely characterized by one irreducible representation of the algebra. This idea becomes central 
later on in section \ref{sec:0Af} when we describe the fermionic matrix model of rectangular matrices, so that we can map the system to free fermions on a sphere.
In section \ref{sec:FQHE} we suggest a possible route to make the particles interact 
so that one can in principle describe a FQHE system on a sphere.  
Iin section \ref{sec:fmm} we describe for completeness the gauged 
fermionic matrix model for square matrices. We show that in this system one has gauged the 
particle-hole symmetry. However, the system still has an edge, which is described by a free chiral fermion on a circle with anti-periodic boundary conditions.
We then conclude.

\section{The gauged $U(N)$ harmonic oscillator and the QHE}\label{sec:matrixQHE}

Let us consider the $U(N)$ matrix quantum mechanics where $X$ is a hermitian 
$N\times N$ matrix, and where $X$ transforms in the adjoint representation of
$U(N)$ by matrix conjugation. We wish to consider the gauged quantum mechanics of 
$X$, where we choose the Lagrangian to be given by
\begin{equation}
L=\int dt \tr\left[\frac12(DX)^2+V(X)\right]
\end{equation}
We will be interested in the potential $V(X)= \frac 12 X^2$ later on,  but for the time
being we will comment on general $V(X)$. There are two ways to solve the system. Choose
the gauge $A=0$, solve the system and impose the gauge constraint. Another way to solve the system is to eliminate the gauge redundancy as much as is possible and solve the system 
in terms of gauge invariant functions of the variables.

As is well known, this second route can be performed if we choose the gauge where 
$X$ is a diagonal matrix with real entries. All Hermitian matrices are conjugate to these matrices by $U(N)$ transformation. Under these conditions the off diagonal components of $X$ and $\dot X$ can be set to zero identically. Therefore the dynamics of the system reduces to the dynamics of the eigenvalues of $X$.
Classically all we have to do is replace a diagonal ans\"atz for $X$ in the Lagrangian to get
the dynamics of the eigenvalues, and a straight forward calculation shows that they are classsically independent of each other.

Quantum mechanically, we have to consider the change of variables from $X$ generic to $X$ diagonal in the wave functions of $X$.
 This produces a change of measure for the eigenvalues of $X$ which is the square of the Van der Monde determinant \cite{BIPZ}.
\begin{equation}
d\mu = \Delta(X)^2 \prod d\lambda_i
\end{equation}
We also have to remember that there is an unbroken symmetry of permutations of the eigenvalues of $X$ which preserves the form of the ans\"atz, and which can be embedded into the gauged $U(N)$ symmetry. This symmetry is gauged, so all of the eigenvalues are 
treated as identical bosons and with measure given by $d\mu$.
This measure dependence can be absorbed in the wave functions of the eigenvalues $\lambda$, $\psi' = \psi \Delta(X)$, with a new measure $d\mu' = \prod d\lambda$. In terms of these wave functions the system describes totally antisymmetric wave functions of the eigenvalues, and 
we have a system of $N$  identical non-interacting fermions in the potential 
$V(\lambda)$. From here, the solvability of the model depends on the 
particular form of the potential $V(\lambda)$. For our purposes $V(\lambda)$ will be the 
harmonic oscillator. The system in the ground state will fill the first $N$ energy levels  of the harmonic oscillator.

We can now equally well consider this system as a set of non-interacting particles in a strong magnetic field. The main idea is that when we reduce a system of particles to the lowest Landau level, the degeneracy of states is captured by a noncommutative plane of magnetic translations. As described in the introduction, this is equivalent to the algebra associated to the phase space of a single quantum variable $X$.  The idea is to identify this noncommutative plane with the phase space of an eigenvalue of $X$. The quantum of area 
is determined by the magnetic field, which will be identified with $\hbar$ after a suitable rescaling of units. Here we think of the 
system as having  a complex matrix $X+iP$, and it's complex conjugate $X-iP$ as a set of conjugate variables.

The degeneracy of states can be broken by a small potential, which is identified with the Hamiltonian for the eigenvalues. This serves to localize the wave functions of the particles in 
the phase space.

Any Hamiltonian function will generically break the degeneracy, but it will not be solvable. Choosing the harmonic oscillator has the benefit of 
producing a rotationally invariant potential with a solvable spectrum. The lowest energy 
state will give a circular droplet whose radius is determined by the number of particles in the droplet. This is due to the Fermi statistics of the eigenvalues. The ground state energy for $N$ particles is exactly $\frac 12 N^2$, and it 
coincides with the ground state energy of the $N^2$ harmonic oscillators in the matrix model.
In this system the energy measures the total angular momentum of the system on 
the plane.

\subsection{Description of the excitations of the system}

So far we have described the system in terms of $N$ free fermions in the harmonic 
oscillator potential, and we have calculated the energy of the ground state.

We now want to describe all excited states of the system. The mathematics of this setup have been recently been reviewed in a work of the author in \cite{Beren}, see also \cite{BKR}.
The complete set of wave functions can be given by 
a Slater determinant of wave functions of the Harmonic oscillator. These wave functions 
are labeled by their occupation numbers $n_1, \dots, n_N$. They are all different and we can use the permutation symmetry so that 
\begin{equation}
n_1>n_2>\dots >n_N\geq 0\label{eq:const}
\end{equation}
The lowest energy configuration has $n_i=N-i$. We will call these values $n_i^0$. The first eigenvalue is chosen at the top of 
the Fermi sea, and then we go down.

To raise the energy, we need to increase the values of the $n_i$ so that they keep satisfying the constraint \ref{eq:const}. If we introduce the quantities $\tilde n_i = n_i- n^0_i$ we see that 
we need to have a non-increasing list of integers $\tilde n_i\geq \tilde n_{i+1}\geq n_N\geq 0$, and that the
energy of the system is given by 
\begin{equation}
E_{\{\tilde n_i\}} = E_0+\sum_{i=1}^N \tilde n_i
\end{equation}
To this state we can associate a Young tableaux with up to $N$ rows, where on row $i$ we 
put $\tilde n_i$ boxes. Young tableaux can also be related to the irreducible representations of $U(N)$ built by tensoring multiple copies of the defining representation and we will make this more precise later on.

A second way to describe the spectrum is given by choosing the gauge $A=0$ first. Then we reduce the system to $N^2$ free harmonic oscillators, which can be described by a set of creation and annihilation operators $(a^\dagger)^i_j$ and $a_i^j$ with commutation relations given by
\begin{equation}
[(a^\dagger)^i_j,a^l_m] = -\delta^i_m\delta^l_m
\end{equation}
and the Hamiltonian of the system is $H = \tr(a^\dagger a) +\frac 12 N^2$.
The vacuum is $U(N)$ invariant and given by the state $\vac$, such that
$a^i_j\vac=0$ for all pairs $i,j$. To build excited states with energy $k$ we act with $k$ raising 
operators in the vacuum and form linear combinations of the states so obtained
\begin{equation}
A^{j_1, \dots j_k}_{i_1\dots i_k} (a^\dagger)^{i_1}_{j_1}(a^\dagger)^{i_2}_{j_2}\dots (a^\dagger)^{i_k}_{j_k}\vac
\end{equation}
Now we need to impose the gauge constraint on these states. This boils down to all upper indices being contracted with all lower indices in some order, so that the state is a singlet under $U(N)$ transformations. We can use matrix 
multiplication to write these states as follows
\begin{equation}
|(s_1, n_1), (s_2, n_2)\dots (s_m, n_m)\rangle
= \tr((a^\dagger)^{s_1})^{n_1} \dots \tr((a^\dagger)^{s_m})^{n_m} \vac
\end{equation}
and we can commute these past each other so that $s_1>s_2\dots >s_m$.

To this state we can also associate a Young tableaux, with $n_1$ columns of length
$s_1$, $n_2$ columns of length $s_2$ and so on. This description gives the same counting of states of the eigenvalue description, provided that we consider $s_i\leq N$. 
This constraint can be seen from the fact that the matrix $a^\dagger$ is an $N\times N$ matrix, and therefore $\tr((a^\dagger)^{N+1})$ can be written algebraically in terms of 
lower traces. This follows from the fact that any $N\times N$ matrix satisfies its 
characteristic equation. Counting these as extra states will produce redundancies.

This second basis looks like a Fock space of states with one oscillator per integer $0<i\leq N$, 
namely $\tr((a^\dagger)^i)$, with energy $i$. This basis is not orthogonal however, so 
the Fock space structure is only an approximation. 
It also follows that this basis can not 
coincide with the basis determined before with Slater determinants, because that basis is orthogonal.

In the thermodynamic limit (large $N$), this approximation of a Fock space is very 
good (the failure of orthogonality of states is small, of order $1/N^2$, this is done by following the  't Hooft idea of counting non-planar diagrams \cite{'tH}. ) and these can 
be described as a free field theory of 
collective excitations of the quantum hall droplet. The states described above change the shape
of the droplet. These are the edge states of the droplet\cite{Stschur}. The oscillator $i$ can be interpreted 
as a wave on the edge of the 
quantum Hall droplet with $i$ units of angular momentum.
This coincides exactly the spectrum of a relativistic  chiral boson on a circle with periodic boundary conditions.

We need a way to relate these two descriptions of the states of the system. 
This is provided by the identification of states with energy $k$ and irreducible representations of $U(N)$ with $k$ boxes. The idea is that we can make a new basis of states by thinking of the matrix $a^\dagger$ as a matrix in $GL(N,\BC)$. Given an irreducible 
representation of $U(N)$ with $N$ boxes, we can elevate it to an irreducible representation of $GL(N,\BC)$ with $k$ boxes. This proceeds by decomposing the tensor product of $k$ fundamentals 
$V^{\otimes k}$ into irreducibles. A group element $g$ acts on $V\otimes V\dots V$ as 
$g\otimes g\otimes g\dots g$. We then project this action onto the irreducibles of the
tensor product by taking suitably symmetrized tensors in the $V$, and we get the action of the matrix $g$ on the given irreducible 
representation.

The character of $g$ in a representation $R$ is given by the trace of $g$ in the given representation, $\chi_R(g) = tr(g)_R$ and it is gauge invariant. Therefore we can make a list of states 
based on the irreducible representations of $U(N)$, by taking the combination $\chi_R(a^\dagger)$. These are the Schur polynomials. Many details of these computations can be found in \cite{CJR}

For example, consider the symmetric representation with two boxes. 
Then 
\begin{equation}
(a^\dagger_S)^{i_1,i_2}_{j_1,j_2}= 
\frac{1}{2}((a^\dagger)^{i_1}_{j_1}(a^\dagger)^{i_2}_{j_2}+(i_1\leftrightarrow i_2)
\end{equation}
And
\begin{equation}
\chi_S(a^\dagger)= \frac12(\tr(a^\dagger)^2+\tr((a^\dagger)^2)
\end{equation}
Similarly for the antisymmetric we get
\begin{equation}
\chi_A(a^\dagger)= \frac12(\tr(a^\dagger)^2-\tr((a^\dagger)^2)
\end{equation}
Notice that the tensor product representation of the two fundamentals has the correct answer 
$\chi_{F\otimes F}= \chi_{S\oplus A}(a^\dagger)=(\chi_F)^2$.
This basis of states coincides with the eigenvalue basis  of Slater determinants that we described first.

\begin{figure}[ht]
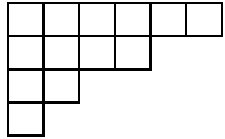

\begin{equation*}
\begin{Young} & & & & & \cr
& & & \cr
& \cr \cr
\end{Young}
\end{equation*}\caption{Young tableaux describing the state $|6,4,2,1\rangle$, $\tilde n_1 =6, \tilde n_2=4, \tilde n_3=2, \tilde n_4 =1, \tilde n_k=0 \ 
\hbox{ for all }\ k>4
$}
\end{figure}

In particular, we can consider states where we take one particle off the Fermi sea and excite it by a large amount. This means we need to take $\tilde n_1$ large, and $\tilde n_k=0$ for all
$k>1$.
These are described by choosing a long row in the Young tableaux, and nothing else: these are totally symmetric representations. Similarly, a hole can be created by taking 
a totally antisymmetric representation of the group $U(N)$.

\begin{figure}[ht]
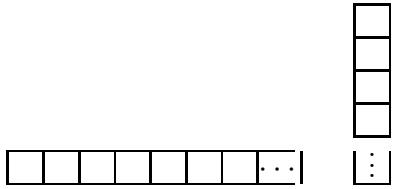

\begin{equation*}
\begin{Young} & & & & & & & $\cdots$\cr
\end{Young}\quad \quad \begin{Young} \cr \cr \cr \cr \vdots \cr
\end{Young}
\end{equation*}\caption{Young tableaux describing a single particle and a single hole state}
\end{figure}

 When we are not too far from the Fermi sea, there is a 
particle-hole symmetry. This symmetry can be understood by noticing that a Young tableaux can be flipped about the diagonal to produce a new Young tableaux. However, if the Young tableaux is sufficiently wide, then when we flip it,  it will no longer be an allowed tableaux for a $U(N)$ representation.

Also if we look at the droplet system as made of two non-mixing liquid system, we have a finite droplet of particles, and an 
infinite droplet of holes.
The symmetry between particles and holes is clearly broken.

The problem we are concerned with in this paper is to find a realization of the quantum hall droplet which has this symmetry manifest in the description of the system, and where it does not appear only after we have solved the system.

Making the symmetry between holes and particles more manifest requires to have either both infinite numbers of holes and particles, or 
finitely many of each. The first option can be realized by the $\hat c=1$ matrix model. This is described by a matrix model with potential given by $-x^2/2$. The Fermi sea  gives a phase space contour determined by a hyperbola centered around zero, and we get either two droplets of particle fluid separated by a thin bridge of hole fluid, or we can get two droplets of hole fluid separated by a thin
bridge of particle fluid. However, the microscopic description of the system is 
given by either particle wave functions or hole wave functions, but they do not appear on the same footing.  Particles are treated as eigenvalues of the matrix, while holes appear when the classical particles (eigenvalues) become quantized fermions, and each fermion occupies a finite area 
on the eigenvalue phase space.

The second option, where we have finitely many particles and holes filling space requires a change in 
topology of the system. This gives us a  phase space of finite volume, so 
that the degeneracy of the Landau levels is finite, and we will only be able to access a Hilbert space of 
states of finite dimension. We will see that the matrix model we propose will be realizing the second option of possibilities.

\section{The type 0A harmonic oscillator matrix model.}\label{sec:0A}

Let us now consider an orbifold of the matrix model we discussed in the 
previous section, the so called type 0A matrix model \cite{DKKMMS}, but we will consider it in the harmonic oscillator case, ass opposed to in the $c=1$ matrix model. The construction of the orbifold is performed by following the ideas in \cite{DM}, and basically leads to a matrix model for a pair of rectangular matrices.
This will 
provide us with some techniques to deal with the fermionic matrix model we will introduce 
later on, and the physical interpretation does not change with respect to the previous discussion very much. 

 The idea  is to orbifold the harmonic oscillator matrix model by the $\BZ_2$ action $x\to -x$. 
This produces a quiver diagram theory with gauge group $U(N)\times U(N+M)$, and $X$ is split into two 
matrices that transform as the $(N, \bar(N+M))$ and the $(N+M,\bar M)$ representations of the group \cite{DM}.
These are complex conjugate to each other, so we can obtain a hermitian matrix model by 
thinking of  a matrix $X$ as 
\begin{equation}
X\sim \begin{pmatrix}0& X_1\\
X_2&0
\end{pmatrix}
\end{equation}
and imposing $X=X^\dagger$.

We can get a complex matrix model with two rectangular matrices of size 
$N\times (M+N)$ and $(N+M)\times M$ respectively, together with their adjoints as their 
conjugate variables in phase space by adding the momenta conjugate to $X$. 
We will call these variables $u,w$, 
and their duals $\bar u,\bar w$.

We can again describe the system in terms of eigenvalues of the composite matrix 
$X_1 X_2= \diag(\lambda^2_i)$ which is positive definite as $X_2= X_1^\dagger$,
 so that we can pick a gauge
where the gauge group is broken down to a diagonal $U(1)^N\times U(M)$, 
\begin{equation}
X_1\sim\begin{pmatrix}
\lambda_1 & 0 &0&\dots\\
\vdots &\ddots &0&\dots\\
0&\dots&\lambda_N& \dots
\end{pmatrix}, 
X_2\sim \begin{pmatrix}
\lambda_1 & \dots &0\\
\vdots &\ddots &\dots\\
0&\dots&\lambda_N\\
0&\ddots&\vdots
\end{pmatrix}
\end{equation}
Again, permutations of the eigenvalues can be embedded in the gauge group, so we are left with symmetric functions of the 
$\lambda_i^2$. There is also a Van der Monde-like determinant (the volume of the 
gauge orbit) which leads to a measure of the form
\begin{equation}
d\mu \sim \prod \lambda_i^{2M} d\lambda_i \prod_{i<j}(\lambda_i^2-\lambda_j^2)^2
\end{equation}
this has been calculated explicitly in \cite{Morris, AMP, DKKMMS} and see also \cite{DiF}.
The integration region for the measure is given by $\lambda_i\geq 0$, as there is no distinction between $\lambda_i$ and $-\lambda_i$ (these are identified by the $\BZ_2$ orbifold action))

After absorbing the square root of the measure in the wave functions, we get antisymmetric wave functions of the 
$\lambda_i^2$, and the effective quantum mechanical system gives rise to the following Hamiltonian
\begin{equation}
H= \frac 12\sum p^2_i+\lambda_i^2+(M^2-1/4)/\lambda_i^2
\end{equation}
with the restriction that $\lambda_i\geq 0$. This has been conjectured to describe a string theory in $AdS_2$ \cite{St}, and we borrow freely some facts from that paper.

This system has an $SL(2)$ algebra for each eigenvalue. Define 
\begin{eqnarray}
D&=&\frac12(\lambda_ip_i
+p_i\lambda_i) = \lambda_i\partial_{\lambda_i}\\ P&=&\frac12(p_i^2+(M^2-1/4) \lambda_i^{-2})\\
K&=&\frac 12 \lambda_i^2
\end{eqnarray}
And it is easy to see that these three operators satisfy an $SL(2,R)$ algebra. 
A different basis for the algebra is provided by
\begin{equation}
L_0=H= \frac{P+K}2, L_+= \frac12(P-K-iD) , L_-= \frac{P-K+iD}2
\end{equation}
This algebra is similar to the $SL(2,\BR)$ algebra of a single harmonic oscillator, with generators $a^\dagger a+1/2$, $(a^\dagger)^2$ and $a^2$. 

It can actually be shown that this is a spectrum generating algebra for the eigenvalues 
of the single particle Hamiltonian. Namely, given a lowest weight state state with 
$L_-\vac=0$,  there is a unique wave function of $\lambda$ which is $L^2$ 
normalizable and satisfies this property. This function 
is 
\begin{equation}
f(\lambda)= \exp(-\lambda^2/2) \lambda^{M+1/2}
\end{equation}
and has energy $M+1$.
Acting $k$ times with $L_+$ we can raise the energy by $2k$ units, and we also make
wave functions $\psi(\lambda^2)\sim \lambda^{1/2+M}\exp(-\lambda^2/2)
g(\lambda^2)$ where $g$ is a polynomial of degree $k$.

This representation of $SL(2,R)$ has lowest weight with spin $M+1/2$, and we will call this the $V_{-M/2-1/2}$ representation. \footnote{The spectrum of the Harmonic oscillator gives two representations of the $SL(2)$ algebra of spin $-1/2$ and $-3/2$}.

The ground state of the system for $N$ eigenvalues will have energy 
\begin{equation}
\sum_{i=0}^{N-1}
(M+1+2i)= MN +N^2
\end{equation} 
By addition of angular momentum , the  $SL(2,\BR)$ algebra of the individual eigenvalues becomes an $SL(2,\BR)$ algebra acting on the Hilbert space of multi-particle states.
The wave functions then belong to the antisymmetric tensor product representation
\begin{equation}
\Lambda^N V_{-M-1/2}
\end{equation}
which contains no $SL(2,\BR)$ singlet. \footnote{Remember that 
$V_{\alpha} \otimes V_{\beta} \sim \oplus_{n=1}^\infty V_{\alpha+\beta -n}$ which for $\alpha,\beta$ negative integers never contains a $V_0$}
For $N=2$ we get for example
\begin{equation}
\Lambda^2 V_{-M-1/2} = \oplus_{n=1}^\infty V_{-2M-2n}
\end{equation}

Again, since the spectrum of states for a single eigenvalue is evenly spaced, we can describe the system in terms of 
a Young tableaux with columns of length less than or equal to $N$, where the length of the rows indicates how much energy over the ground state 
we have put into each eigenvalue, starting from the top of the Fermi sea.
The system can also be described in terms of bosonic eigenvalues. This follow from the
identity of representations
\begin{equation}
\Lambda^N V_{-M-1/2} = S^N V_{-M-N/2}
\end{equation}

We can again think of the system as describing a quantum hall droplet. Due to the 
$SL(2,\BR)$ symmetry we can consider it as giving the holomorphic 
quantization of wave functions on the Poincare disc (or under a conformal transformation by the upper half plane), which is  
$SL(2,\BR)/U(1)$.

Now we can repeat the procedure for the system as we did in the last section, and go again to the gauge $A=0$. For each component of 
$X_{1,2}$ we get an harmonic oscillator. The creation operator and annihilation 
operators are given by $u,\bar u\sim a^\dagger, a$ and $v,\bar v\sim b^\dagger, b$.

The $SL(2,\BR)$ algebra is given by \begin{eqnarray}
L_0&=& H= \tr(a^\dagger a+b^\dagger b)+M(M+N)\\ 
L_+&=& \tr(a^\dagger b^\dagger)\\
L_-&=& \tr(ab)\end{eqnarray}

Again, we get a description in terms of waves on the edge of the droplet by 
considering traces $\tr((a^\dagger b^\dagger)^n)$, each of which represents one 
quantum with angular momentum $n$ on the edge, and energy $2n$. 
This description becomes accurate in the thermodynamic limit of large $N$.
Notice that the normalization of $L_0$ and $H$ differ by a factor of $2$ from the standard $SL(2)$ normalization. 

We can also describe the system in terms of representations of $U(N)$ with $k$ boxes. This works very similar to the previous discussion in the ordinary matrix model. 
The only difference is that we need to take the composite matrix 
$a^\dagger b^\dagger$ as the $N\times N$ matrix. Here, we get representations of 
$U(N)$ and $U(N+M)$ to work with. We take properly symmetrized 
products of $(a^\dagger)^{i_k}_{j_k}$ in the upper $U(N)$ indices characterized by a given Young tableaux. 
The fact that these operators all commute with each other means that whatever symmetry the 
upper indices have, it is mirrored in the lower indices. Therefore the Young tableaux for $U(N)$ and $U(N+M)$ are identical. The orthogonality of the basis thus constructed is 
fairly easy to prove. This basis should coincide with the eigenvalue basis.

The lesson we should learn is that traces always represent edge states of the system, and
that Schur functions (characters of composite operator on irreducible representations) represent Slater determinant wave functions.
Notice that again in this case we have an infinite volume of holes, and a finite volume of particles. Notice also that in the description of the character basis there is a correlation between Young tableaux for $U(N)$ and
young tableaux for $U(N+M))$, they are indeed identical, so that only those tableaux of $U(N+M)$ are allowed which are also tableaux for $U(N)$. In this sense, exchanging 
$N$ by $N+M$ does not have any effect on the Young tableaux description of the states.
This does not exchange particle and hole like tableaux. This is why we said in the introduction that this model does not work to describe particles and holes in a symmetric way.

\section{A fermionic matrix quantum mechanics with $U(N)\times U(M)$ symmetry}
\label{sec:0Af}

Now we are ready to describe another matrix model, which is the main result of this paper.
 The basic idea is very simple: 
consider the type 0A harmonic oscillator matrix model, except that we use fermionic 
oscillators instead of bosonic ones. Thus we have a system with two 
rectangular matrices of fermionic creation operators of size $N\times M$ and 
$M\times N$, and we impose a $U(N)\times U(M)$ invariance on the allowed wave 
functions. Ordinary matrix models for fermionic variables have been studied in \cite{MZ,SS}
and they have similar properties to ordinary matrix models. Here we are studying the fermionic
quantum mechanics for the matrix harmonic oscillator with rectangular matrices.

The first thing we should notice is that the fact that we have fermions makes various 
changes to the system. First, there are only finitely many degrees of freedom (without using the gauge invariance there are $2^{2NM}$ states). The second point is that in some sense fermionic variables can only be interpreted in terms of operator algebras, but they can not be thought of as numbers. Because of this, it is not possible to diagonalize an $m\times m$ 
matrix of fermions to discover the eigenvalues. The reason for this is that as operators, the entries of the matrices do not commute, they anti-commute. Therefore they can not be diagonalized 
simultaneously as operators to obtain a matrix of c-numbers on states which can be diagonalized. This 
makes some aspects of the description of the system a little bit awkward, because we have lost part of the semiclassical description.

We still have the second alternative of using the gauge $A=0$ and writing gauge 
invariant wave functions by taking traces. This gives us the same behavior as
the edge of a quantum hall droplet: we obtain one oscillator 
of the edge per integer $n$ in the thermodynamic limit $N\sim M$ large. 
The creation operator for such an edge state of momentum $n$ is given by $\tr((a^\dagger b^\dagger)^n)$. Notice that these are bosons, because they are made out of an even number of fermionic operators, thus these operators commute. Again one can show that in the thermodynamic limit there is a  similarity to a Fock space of bosons made out of
these states, and that they are approximately orthogonal to leading  order in $1/NM$, so long as we keep the energy finite and not scaling in the limit. We do see that the system is describing an edge of some type of quantum hall droplet. Our final purpose is  to investigate this in more detail.
In this paper we are interested in the case $N,M$ finite, and not in the thermodynamic limit of the system itself.

Here, the ground state of the system has energy $-NM$. The negative number is 
the standard fact that free fermions contribute oppositely to the zero point energy than free bosons.

Since these states built out of traces are not orthogonal to each other, they are not a good basis for wave functions of the system. As we have seen, there is a second basis of orthogonal 
states which correspond to a basis of eigenvalues. These are obtained using characters of
the groups $U(N)$ and $U(M)$ associated to different irreducibles of $U(N)$ and $U(M)$. We will proceed with this description now.

\subsection{Description of the spectrum in terms of Young tableaux.}

Let us consider $\chi_R(a^\dagger b^\dagger)$ for $R$ a young tableaux of $U(N)$.
The Fermi statistics of the $(a^\dagger)^i_j$ show that if we have two upper indices which 
are symmetrized, then the two lower indices are anti-symmetrized. Similarly if 
the upper indices are antisymmetrized, then the lower indices are symmetrized. This simple observation is at the heart of our claims about the properties of this system.

\begin{figure}[ht]
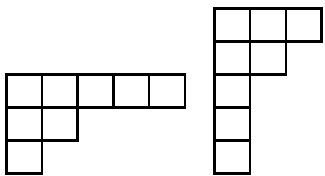

\begin{equation*}
\begin{Young}
& & & & \cr
& \cr
\cr
\end{Young} \quad \begin{Young} & & \cr
&\cr
\cr\cr\cr
\end{Young}
\end{equation*}
\caption{Flipping the tableaux: upper $U(N)$ indices and lower $U(M)$ indices of tensors made of products 
of $a^\dagger$ are related by Fermi statistics by flipping the tableaux through the diagonal.\label{fig:flip}}
\end{figure}

This means that the representations $R$ of $U(N)$ and  $\tilde R$ of $U(M)$ are correlated
\begin{equation}
\chi_R(a^\dagger b^\dagger) = \chi_{\tilde R} (b^\dagger a^\dagger) 
\end{equation}
where we have to consider two Young tableaux which are mirror images under the reflection on 
the diagonal. Notice that above $a^\dagger b^\dagger$ and $b^\dagger a^\dagger$ are matrices of size $N\times N$ and $M\times M$ respectively. This is presented in figure \ref{fig:flip}.

In particular, the restrictions on the tableaux from being allowed both in $U(N)$ and 
$U(M)$ tell us that the tableaux for $U(N)$ have columns of length less than $N$ and
rows of length less than $M$, while the ones for $U(M)$ have columns of length less than 
$M$ and rows of length less than $N$. Each box adds two units of energy to the system.
The description of the spectrum is essentially symmetric in the exchange $M\leftrightarrow N$, except that we also have to flip the Young tableaux along the diagonal. 

In particular, the state with the maximum number of boxes allowed has energy 
$(-NM)+2NM=NM$, which is the highest energy state of the ungauged  matrix 
model.

The system has an underlying $SU(2)$ algebra, similar to the $SL(2,R)$ algebra for the type 0A matrix model. This is generated by the Hamiltonian, $H\sim L_0$, $L_+=\tr(a^\dagger b^\dagger)$, and $L_-=\tr(ab)$. The difference between $SU(2)$ and $SL(2,\BR)$ is the sign with which $L_0$ appears on the commutator of 
$L_+$ and $L_-$. Notice that $(a^\dagger, b)$ and $(- b^\dagger, a)$ form two doublets of operators under this $SU(2)$ symmetry.

The Hamiltonian is $L_0$, so it breaks the $SU(2)$ symmetry of the system. However, 
the Cassimir operator $L^2$ commutes with $H$, so it is a good quantum number. 
Thus we can describe the spectrum in terms of the representation theory of $SU(2)$.
If we want to associate this symmetry with a two-dimensional topology, it is naturally the 
isometry group of a 2-sphere, so it suggests that the associated quantum hall droplet should be a state  on a sphere.

We can easily see that the system is being given by $N$ bosons in the spin $M/2$ representation of $SU(2)$ by counting the degeneracies of $L_0$. This is the same as having $M$ bosons in the spin $N/2$ representation. Roughly speaking, each of the bosons is given by the rows of the Young tableaux with respect to $U(N)$.

We write this Hilbert space as follows
\begin{equation}
S^N(V_{M/2}) \sim S^M(V_{N/2})
\end{equation}

In this description we can think of the system as being given $N$ free charged bosons on the lowest Landau level of a sphere with magnetic monopole of strength $M$, in the 
presence of a small amount of gravity  (or an electric field pointed on 
the same direction: this is the source of the potential proportional to $L_0$) 
that makes them settle to the bottom of the sphere.
This description is dual to $M$ free bosons on the lowest Landau level on a sphere with a magnetic monopole field of strength $N$ in the presence of gravity. In these two decriptions, we see that we have a duality where we exchange number of particles with flux on a two
sphere. 

We should point out that in the description of the original matrix model we 
get that  $N$ and $M$ appear symmetrically. This symmetry is broken depending on how we choose to interpret the system: either $N$ or $M$ bosons on geometries with different monopole backgrounds. 

The total number $n$ of states is 
\begin{equation}
n = {N+M \choose M} = {N+M\choose N}
\end{equation}
which also suggests a description in terms of $N$ or $M$ 
fermions with $N+M$ states available to 
each. This is the description we will look for now.

\subsection{Fermionization of the  ``eigenvalues" and $SU(2)$.}

So far we have obtained a description of the Hilbert space in
terms of identical bosons. We can fermionize this description. The idea is to 
think of the system as $N$ fermions on a sphere with monopole background 
determined by the number of states available per fermion. To get $N+M$ states from which 
we occupy $N$,  we 
need to start with the spin $(N+M-1)/2$ representation of $SU(2)$. 
This corresponds to a monopole background  on a two sphere 
of charge $N+M$. Remember that we can interpret the 2-sphere as the homogeneous space $SU(2)/U(1)$. This is analogous to the discussion of the type IIA bosonic matrix model 
where we had fermions on $SL(2,\BR)/U(1)$.
 We also have a small electric field breaking the degeneracy of the states.

It is easy to show that the representation theory of $SU(2)$ found before is given by
\begin{equation}
S^N(V_{M/2}) \sim \Lambda^N(V_{(M+N-1)/2)})\sim \Lambda^M(V_{(M+N-1)/2)})
\end{equation}

Because of Fermi statistics, the $N$ fermions will fill the bottom of the 
$S^2$ sphere, leaving $M$ holes at the top as we have a total of 
$N+M$ states available to each Fermion. So now we see that $N$ and $M$ can be
interpreted as the number or particles and holes  on the sphere respectively, and we have a droplet of quantum hall liquid for particles at the bottom and for holes at the top.
The interface between the two liquids will correspond to the edge of the droplet. We already have the candidate states that describe the changes in shape for the droplet, given by traces of
the powers of  $a^\dagger b^\dagger$. These don't distinguish between $U(N)$ and $U(M)$ symmetry because of the cyclic property of the trace.

The energy of the ground state will be (if we count the particle states)
\begin{equation}
\sum_{i=1}^{N} (-M-N+2i+1)=-(M+N)N+N^2= -MN
\end{equation}
which coincides with the matrix model. This is just the $SU(2)$ angular momentum of the lowest weight state in the antisymmetric tensor product.

Now, the next step is to decide how to interpret the particle and hole excitations in the 
system. Following our previous discussion of the one matrix model 
in terms of Young tableaux, we create particles by considering symmetric representations of $U(N)$, and holes by considering antisymmetric representations of $U(N)$, these are equivalent to antisymmetric and symmetric representations of $U(M)$ respectively, from the pairing of Young tableaux. Again, we see that the exchange $M\leftrightarrow N$ exchanges the notions of 'holes' and 'particles'.

The first ones should build one large eigenvalue for the $N\times N$ matrix 
$a^\dagger a$, while the second ones build a large eigenvalue for the 
matrix $b^\dagger b$.

 It should be clear by now that we have a matrix model which describes a quantum hall 
 droplet on the topology of the sphere, and which is also in the presence of gravity or an electric field. In the thermodynamic limit it has an exact 
 $c=1$ CFT on the edge. Moreover in the matrix model the particles and holes ($N$ and $M$) appear symmetrically in terms of the matrix degrees of freedom.
 
The coordinates of the particles and holes themselves (as considered by the eigenvalues of 
$a^\dagger a$ and $b^\dagger b$) appear as different 
composite fields of  gauge variant fields, which are not directly observable. This is what makes possible the symmetry between particles and holes to be present in this formulation of the model.

This description is in terms of free fermions: $N$ particles or $M$ holes, and we can go back and forth between these by the identification of the following naturally dual vector spaces
\begin{equation}
\Lambda^N V_{\frac{M+N-1}2} = \Lambda^N V^*_{\frac{M+N-1}2}
\end{equation}
where we use the Hilbert space norm to identify $V$ and $V^*$.
Now that we have identified a matrix model which has all the states to describe a finite number of fermions on a finite geometry we can perturb the Hamiltonian to obtain other 
interesting models, which can include interactions between the fermions.

\section{Towards the FQHE: adding interactions.}\label{sec:FQHE}

The fractional quantum hall effect can be obtained by considering fermions which 
fill a fraction of a Landau level and which have repulsive interactions \cite{Laughlin}.
 We also have to go 
to the thermodynamic limit so that the number of particles and holes both scale the same way as we take the number of particles to infinity, keeping $N/M$ finite and rational. In this paper we so far have ignored the details of the large $N,M$ limit, so we will try to provide a 
method for studying the system at finite values of  $N,M$. For the topology of a sphere, the fractional quantum hall state has also been considered in \cite{Haldane}

In our case we already have described a system of free fermions, so now we need to 
add interactions between them by perturbing the Hamiltonian of the model. Since we have 
the correct Hilbert space to describe all wave functions in the corresponding Landau level, 
there exists a perturbation that 
will produce the desired effect on all of the states. What is not clear, is how simply this 
perturbation is described in terms of the natural variables we are working with (the 
$a,b$ fermionic oscillators), and how much of the details of the FQHE depend on 
the exact form of the Hamiltonian.

We would want to perturb the system by adding as few terms as possible, and we definitely 
want to do it in 
such a way that the terms that we add are $SU(2)$ invariant and that they involve only two particles at a time. This last part is where this formulation might become cumbersome. 
The only term which would break the $SU(2)$ symmetry is the unperturbed Hamiltonian, and this will favor the state with largest negative eigenvalue of $L_0$.

The requirement of invariance under $SU(2)$ transformations places various constraints 
on the perturbation. In particular, it has to commute with $L_0$, so the
additional term in the Hamiltonian and $L_0$ are always mutually 
diagonalizable.  The hamiltonian will then mix states with the same number of boxes (and which also belong to the same eigenvalues of $L^2$), and so long as the perturbation is small, the lowest energy state will be a state with a small number 
of boxes (compared to $N_1$ and $N_2$), 
so it can be analyzed as a small perturbation of the edge of the quantum hall 
droplet. \footnote{This procedure  would not realize the fractional quantum hall state as the lowest lying state, which has $L^2=0$ and energy $MN/2$ over the vacuum in the 
free particle limit.}

This also means that in the matrix basis, the perturbation will always have the same 
number of raising and lowering operators.

For the description to be simple in terms of the matrix variables, we would want to 
have a polynomial with few terms in the fermionic fields. 
The simplest terms that we can add involve 
two raising and two lowering operators.

In the description we had above, all states with same values of $L_0$ were degenerate in energy, irrespective of their total $SU(2)$ quantum number. The simplest operator we can 
add that breaks this degeneracy is a term proportional to the total angular momentum 
\begin{equation}
\delta H =\alpha L^2 = \alpha( L_+L_- +L_0^2+2L_0)
\end{equation}
We also have the freedom to add a constant term to $\delta H$ so 
that  the  energy of the lowest state we had before is not altered. It is easy to see that 
this perturbation of the Hamiltonian has exactly two raising and two lowering operators.

This perturbation is of double trace type, as each of $L_+,L_-$ and $L_0$ 
is of single trace type. Single trace operators for diagonal 
matrices can't produce interactions between the eigenvalues (this is a statement one makes with classical diagonal matrices). Double trace operators produce interactions between pairs of eigenvalues, triple trace operators produce interactions between 
three eigenvalues at a time and so on. In general, for two body interactions we would expect that the Hamiltonian is of double trace type. Double trace deformations of large $N$ systems are also solvable \cite{KH}

The first term we add clearly lifts some of the degeneracies. It removes all degeneracies for 
a system with just two particles or two holes, but in general for higher numbers of particles this will not be enough, as we will have various representations with the same value of $L^2$.
This will be true for any function $f(L^2)$, but the restriction of being double trace basically forces us to consider only the term above.

The next thing we can do is classify all the single trace operators according to their $SU(2)$ quantum numbers. Let us consider a trace with only creation operators. For example
\begin{equation}
\tr((a^\dagger b^\dagger)^n)\label{eq:hweight}
\end{equation}
This state is the highest weight state of a spin $n$ representation of $SU(2)$. Acting with lowering operators changes some of the $a^\dagger$ and $b^\dagger$ for $b, a$ 
respectively.

Now, since we have matrix valued variables, the order of operators inside the trace matters.
In principle this means that the two operators 
\begin{equation}
\tr(a^\dagger a a^\dagger b^\dagger), \quad \tr(a^\dagger b^\dagger b b^\dagger)
\end{equation}
would be linearly independent.

However, we are imposibg the gauge constraint on the system, which reads
\begin{eqnarray}
:a^\dagger\cdot  a + b\cdot b^\dagger: &=& 0\\
:a \cdot a^\dagger + b^\dagger \cdot b: &=& 0
\end{eqnarray}
The multiplication above is matrix multiplication. The operators written above are the generators of the $U(N)\times U(M)$ symmetry. They are normal ordered, but we keep the matrix order as written above.

On physical states, the above operators vanish. This means that when we find these operators inside a trace, we can use these relations to replace one matrix operator by another one.

We see this way that different orderings of the letters $a, b$ don't really matter too much, and the invariant of the collection of such operators is the length of the word and the spin $L_0$.
The length tells us that the word is obtained from the highest weight state \ref{eq:hweight} by commutators with $L_-$.

In essence, the total set of single trace operators that we can consider fall into the representations 
\begin{equation}
0\oplus 1\oplus 2\oplus \dots \oplus \tilde N\label{eq:reps}
\end{equation}
where $\tilde N$ is the smallest of $M,N$.
After we get to $\tr((a^\dagger b^\dagger)^{\tilde N})$ there are no more algebraically independent traces that we can consider.

From here, the double trace operators will come from considering the singlets in 
the $k\otimes k$ representation of $SU(2)$, one coefficient for each of the 
representations above, in equation \ref{eq:reps}.

Our hamiltonian will contain one coefficient per representation of $SU(2)$. These should be related to the potentials that Haldane\cite{Haldane} used written in terms of 
$L_i\cdot L_j$ for different particles.

It would be interesting to determine how many of these coefficients should be needed to
realize a fractional quantum hall state. The observation that one expects a fractional quantum hall system when the interactions between the particles are short range and repulsive should indicate that we need to go to high spin for the operators to mimic this effect. The terms with low spin are related to the long range potentials between the particles.

\section{A system with a $c=1/2$ edge}\label{sec:fmm}

We have so far described a matrix  model which represents a quantum hall droplet on a sphere. 
This model is an "orbifold" by a $\BZ_2$ action, just like the 0A matrix model was an orbifold by a $\BZ_2$ action. For completeness, we should describe the associated fermionic system without orbifolding. THis is just the gauged fermionic quantum mechanics, based on one fermionic matrix oscillator pair $f, f^\dagger$.

The discussion is not too distinct from the our original fermionic matrix model. 
The spectrum does not allow for fermionic eigenvalues, so we also need to be careful to describe the spectrum correctly.

From the point of view of traces, we get the states
\begin{equation}
c_n= \tr((f^\dagger)^n)
\end{equation}
Notice that by the cyclic property of the trace, $c_n= (-)^{n+1} c_{n}$, so that the only values of $n$ allowed are the odd integers. One also uses  this property to show that 
$\{c_{n},c_{m}\}=0$, so that these states are fermions.

There is one fermion mode per odd integer. This is similar with the mode structure of a chiral fermion on a circle with antiperiodic boundary conditions. That system 
has a $c=1/2$ central charge. We conclude that this fermionic matrix model is describing 
an 'edge' with a $c=1/2$ central charge. 
It is not usual that one would find such a system in a FQHE, as the central charge is usually greater than or equal to one to accomodate for the possibility of adding and removing charge from the system. This leads to a $U(1)$ current algebra that measures this charge and has 
central charge $c=1$, see \cite{Wen,Wen2,CDTZ} for example, and more recently \cite{BCR}. There can be additional degrees of freedom which might raise the central charge even further (for example, the Pfaffian state of Moore and Read  has this property \cite{MR}).

\begin{figure}[ht]
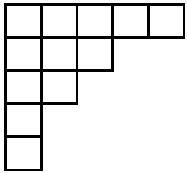

$$\begin{Young}
& & & & \cr
& & \cr
 & \cr
 \cr \cr
\end{Young}
$$ \caption{An allowed Young tableaux for the fermionic matrix model\label{fig:allow}}
\end{figure}

From the point of view of Young tableaux, we need to remember that for fermions, as described previously, the upper and lower indices in a Young tableaux transform in representations that are
related by flipping along the diagonal.  To make a gauge invariant state, the upper and lower indices should correspond to the same tableaux, son only those tableaux that are symmetric along the diagonal are allowed, this is shown in figure \ref{fig:allow}.

Notice that the counting of states coincides with the counting via traces. This is because one can count the energy of a fermion by a hook centered on the diagonal of the tableaux. In the example in figure
\ref{fig:allow}, one has two hooks of length $9$ and $3$ respectively. It is easy to see also that no pair of these hooks have the same length in any tableaux.
From the point of view of particles and holes (rows and columns of the tableaux), the Young tableaux indicate that whenever we give a large amount of energy to a particle we also are giving the same amount of energy to a hole. The system 
is therefore describing correlated particle-hole states. Notice also that the system does not have the $SU(2)$ symmetry anymore either. This is because the generators $L^+$ and $L_-$ from the previous section are not allowed operators.

One can consider a third description of the system, based on a different approach to understanding the fundamental degrees of freedom as fermionic matrices. 
The idea is to remember that 
one can think of differentials as fermions, so here we have a set of differential forms 
which are Lie algebra valued for the Lie algebra of $U(N)$ matrices. If we consider 
the set of $U(N)$  equivariant forms for the Lie algebra, we will be describing the
gauge invariant states. the relations between the Lie algebra and the Lie group itself 
can be used to rethink the problems in terms of the (differential) cohomology groups of the $U(N)$ manifold itself. It is known that $SU(N)$ can be considered as a sphere bundle $S^{2N-1}$ over the group $SU(N-1)$, and that the cohomology of $SU(N)$ is the cohomology of
of the above sphere tensored with the cohomology for $SU(N-1)$. We can proceed by induction 
to get algebraic generators given by a $S^3$, an $S^5$, etc. Here our hamiltonian
is the degree of the differential form over $SU(N)$. To consider $U(N)\sim U(1)\times SU(N)/Z_{N}$, we remember that it 
is essentially a product space, so we also get a generator for degree one from the $U(1)$ circle. Thus the system can also be considered as a topological model on the $U(N)$ group manifold. The counting of states we obtain this way is the same as the one we described previously. There is one fermion oscillator for each odd integer.

The orbifold of this system is obtained by the $\BZ_2$ identification $f\to -f$, and this recovers the quantum hall droplet on the sphere. An orbifold system by an abelian symmetry also has a quantum symmetry that one can orbifold by and recover the original system. 
This acts by exchanging $a\leftrightarrow b$ in the section \ref{sec:0Af}, so that one ends up identifying the two.  One can recover the original formulation by thinking of this as an orbifold of an orbifold, along the lines of \cite{BJL}. Notice also that this particular $\BZ_2$ symmetry is exactly the one that identifies particles with holes. The $\BZ_2$ quantum symmetry does not commute with the $SU(2)$ action however, as it does not identify the doublets properly. This is in contrast to the case of the bosonic matrix model, where the $SL(2,\BR)$ symmetry is present in both the type 0A and type 0B theory.

It would also be interesting to investigate if it is possible to deform the theory by
double trace operators and get exotic states which are somewhat analogous to a fractional quantum hall state: a quantum liquid with a gap, but with no charged excitations.

\section{Conclusion}

We have argued in this paper that a particular  fermionic gauged  matrix quantum mechanics, which describes of a pair of rectangular matrices,   provides a matrix model description of a quantum hall droplet of non-interacting particles. This quantum hall droplet lives on a geometry with the topology of a 
sphere, and the model has a manifest particle-hole symmetry.

The way our system is able to do this is that neither the holes nor the particles are manifest in the model. Rather, it is by looking at the spectrum of the model that one is able to identify particle and hole wave-functions in the description. Indeed, these can be considered as eigenvalues of 
two different matrix-valued composite operators. 

Since we have this particle-hole symmetry manifestly in the system, it is possible to gauge it by performing an 'orbifold of an orbifold construction', as in \cite{BJL}. Doing this we recover a single gauged matrix model for a single fermion.
This gave rise to a quantum droplet with an edge which has a $c=1/2$ CFT, namely, a theory with a free fermion on a circle. It is always interesting to ask if it is possible to find some realization of the periodic boundary condition as well, as in the CFT this would correspond to the description of twist operators.

We have also argued that it is possible to include interactions between the particles, which does not look too complicated and maybe it is enough to describe fractional quantum hall phases in the thermodynamic limit. 

Recently, there has also been a string theory dual proposal for the gauged 
harmonic oscillator system \cite{IMc}, and see also \cite{Beren}. It would not be surprising if it is also possible to find such a dual
string theory for the fermionic oscillator we have described here.

\section*{ Acknowledgments }

I am very grateful to L. Balents, M . Fisher, E. Fradkin, J. McGreevy, I.Klebanov and J. Polchinski for various discussions related to these ideas. I also would like to thank the Banff international research station where some of this work was done. Work supported in part by a DOE outstanding Junior Investigator award, under grant DE-FG02-91ER40618.

 \end{document}